\documentclass[prl,aps,twocolumn,amsmath,amssymb,superscriptaddress]{revtex4-1}

\usepackage{graphicx}
\usepackage{bm}

\begin{document}

\title{Crystal structure of the inversion-breaking metal Cd$_2$Re$_2$O$_7$}
\author{M. R. Norman}
\affiliation{Materials Science Division, Argonne National Laboratory, Argonne, IL  60439, USA}

\date{\today}

\begin{abstract}
Second harmonic generation (SHG) on the pyrochlore metal Cd$_2$Re$_2$O$_7$ indicates the presence of three order
parameters setting in below an inversion breaking transition.
Here, we explore a possible structural explanation and relate it not only to the SHG data,
but also to neutron and x-ray diffraction, where we find that such a structural scenario can explain certain reflection extinctions
observed in single crystal x-ray data.  From this analysis, we suggest future experiments that could be done to resolve this matter.
Finally, we comment on the Landau-violating nature of the inversion breaking transition and its relation to similar
phenomena observed in improper ferroelectrics.
\end{abstract}

%\pacs{78.70.Ck, 75.25.-j, 75.70.Tj, 42.65.-k}

\maketitle

Although most pyrochlore oxides are insulating, Cd$_2$Re$_2$O$_7$ is a notable exception.  It is a good metal and
exhibits three phase transitions, one near 200 K ($T_{s1}$), another near 120 K ($T_{s2}$), and finally a low temperature superconducting
transition ($T_c$) \cite{review}. Structural data support the presence of inversion breaking at $T_{s1}$ from a high temperature cubic
phase (Fd$\bar{3}$m) to a tetragonal phase (I$\bar{4}$m2), whereas $T_{s2}$ is claimed to be a weakly first-order 
transition to I4$_1$22 \cite{hiroi}.  There are several issues, though.  First, x-ray \cite{xray} and neutron \cite{neutron} data indicate
very different atom displacements, these measurements being hampered by the fact that the structural distortion is weak.
Second, single crystal x-ray diffraction \cite{single} find reflection extinction conditions that are not consistent with I$\bar{4}$m2.
Finally, second harmonic generation (SHG) data \cite{petersen} are consistent with  I$\bar{4}$m2, but are not consistent with
its conversion to  I4$_1$22 below $T_{s2}$.

Subsequently, Liang Fu suggested the possibility that a novel spin nematic phase might set in below $T_{s1}$ \cite{fu},
motivated by the fact that large changes occur in the resistivity and susceptibility at $T_{s1}$ despite the weak nature of the
structural distortions.
This it turn motivated new SHG experiments \cite{harter} that supported this picture, in that they found that 
the previously observed SHG signal was secondary in nature in regards to its temperature dependence (that is, it varied
as $T_{s1}-T$).  They found a new (but much smaller) signal that had a primary order parameter dependence ($\sqrt{T_{s1}-T}$).
From the azimuthal and polarization dependence of the SHG signal, the secondary one (as before) was identified as having E$_u$ symmetry relative
to the high temperature cubic phase
(consistent with I$\bar{4}$m2), but the primary one was identified as having T$_{2u}$ symmetry.  Given the known temperature
dependences of the two SHG signals,
this implies the presence of a third-order term in the Landau free energy.  Since a product of three odd parity
order parameters is not allowed, this indicates the presence of a third order parameter of even parity that also has a primary
temperature dependence, identified as T$_{1g}$.  These two primary order parameters were speculated to be associated with
spin nematic order, with the secondary order parameter being a weak structural effect driven by the primary order.
The three together define a trilinear term in the Landau free energy.

In previous work \cite{matteo}, we did a detailed analysis of possible symmetries associated with the SHG data.
We verified the structural origin of the secondary order
parameter, finding that its SHG signature can be expressed in terms of an axial toroidal quadrupole (identified as
well in later work \cite{japan}).  We briefly mentioned a possible structural scenario for all three order parameters before turning
to a possible magnetic one for the primary order parameters based on the known magnetic order observed in the closely
related pyrochlore Cd$_2$Os$_2$O$_7$ \cite{yamaura}.
But to date, there is no evidence from neutrons, Raman, nuclear magnetic resonance, or nuclear quadrupole resonance for 
either magnetic dipolar or spin nematic order.

All of the above motivates the present work, which instead studies the possibility of a structural origin for the primary order
parameters.  In that context, we note the possible relation to Cd$_2$Nb$_2$O$_7$, whose primary structural order parameter
also has T$_{2u}$ symmetry \cite{malcherek}.  We then relate this to previous x-ray and neutron diffraction experiments, and suggest
future experiments that could resolve this matter.  Finally, we comment on the the fact that condensing two primary order parameters
at a second-order phase transition is a strong violation of the Landau theory of phase transitions,
and connect this to equally puzzling data for improper ferroelectrics \cite{bousquet}.

A possible structural scenario can be obtained by looking at all possible group-subgroup relations.  From the Bilbao crystallographic
server \cite{bilbao}, the most likely structural scenario is I$\bar{4}$ as we mentioned before \cite{matteo} (Fig.~1).  The three space groups
that feed into this are I$\bar{4}$m2 (E$_u$), I$\bar{4}$2d (T$_{2u}$) and I4$_1$/a (T$_{1g}$).  The product of these three define a
trilinear term in the Landau free energy as in Ref.~\onlinecite{harter} and summarized in Table I.  By analyzing the atom displacements
relative to the cubic phase consistent with these space groups \cite{isotropy},
one can easily verify that  I$\bar{4}$2d  (T$_{2u}$) gives rise to an $xy$ axial toroidal quadrupole,
and I4$_1$/a (T$_{1g}$) to the $z$ component of an axial toroidal dipole, with I$\bar{4}$m2 (E$_u$) giving rise
to an $x^2-y^2$ axial toroidal quadrupole.
As a reminder, the antisymmetric combination of $r$
(cubic atom positions relative to
the inversion origin) and $d$ (their displacements from $r$ in the distorted phase) is the axial toroidal dipole 
($g \equiv r \times d$, a polar vortex which preserves inversion).  The symmetric
combination of $g$ and $r$ form a pseudoscalar ($r \cdot g$) and five quadrupolar components (i.e., a pseudodeviator), all of which
break inversion.

\begin{figure}
\includegraphics[width=0.75\columnwidth]{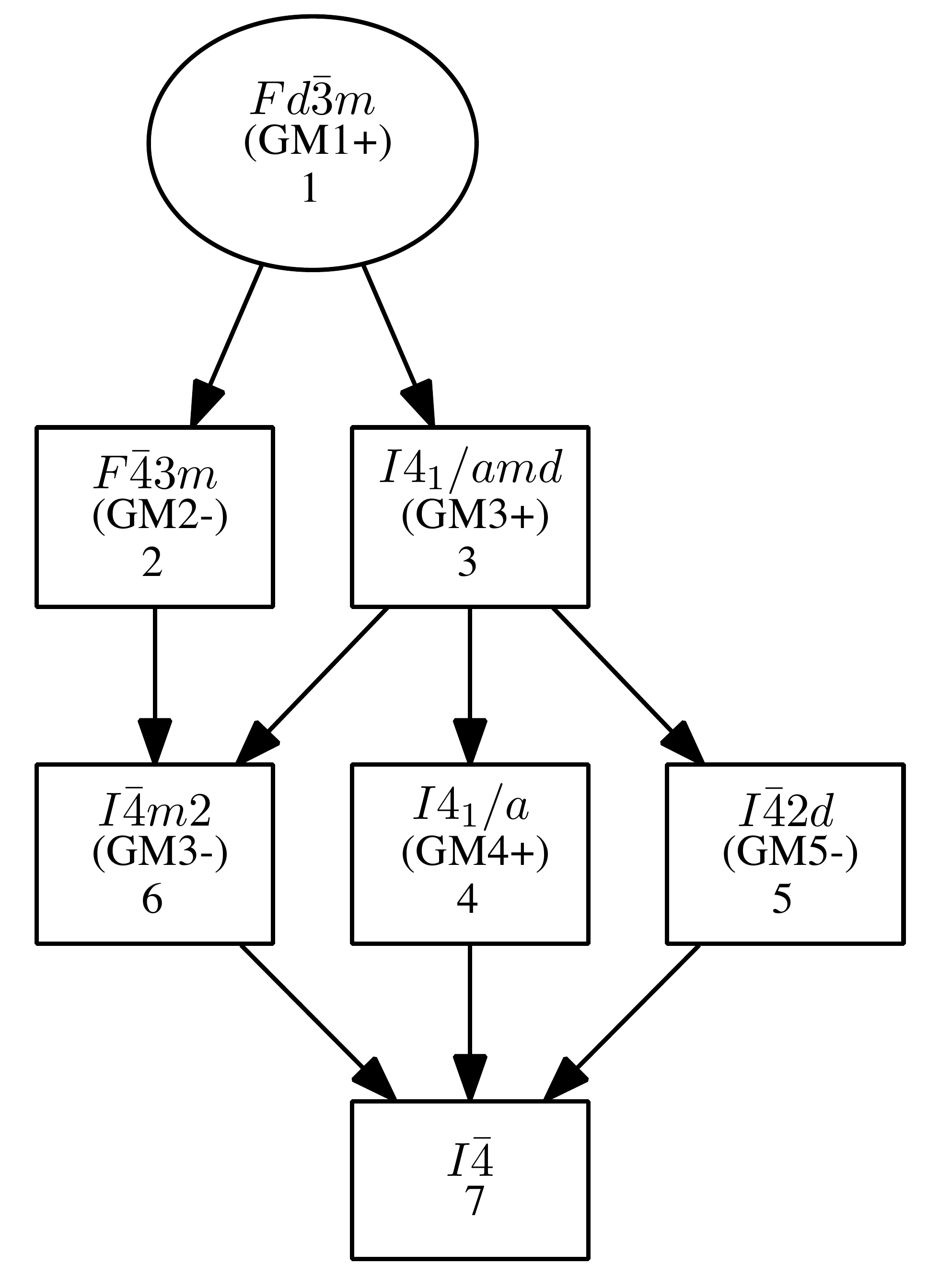}
\caption{Group-subgroup relations leading from Fd$\bar{3}$m to I$\bar{4}$ \cite{bilbao}.  Here, the cubic group representations
are A$_{1g}$ ($\Gamma_1^+$, GM1+),  E$_g$ ($\Gamma_3^+$, GM3+), T$_{1g}$ ($\Gamma_4^+$, GM4+), A$_{2u}$ ($\Gamma_2^-$, GM2-), E$_u$ ($\Gamma_3^-$, GM3-), and T$_{2u}$ ($\Gamma_5^-$, GM5-).}
\label{fig1}
\end{figure}

\begin{table}
\caption{The three order parameters (two primary, one secondary) that define the trilinear term in the Landau free energy in the
two structural scenarios (with the space groups associated with them in parentheses).
Note the order parameters for the first scenario have the same symmetries as those defined in the
electronic scenario proposed by Harter {\it et al.}~\cite{harter}.}
\begin{ruledtabular}
\begin{tabular}{lrrr}
space group & primary-odd & primary-even & secondary-odd \\
\hline
I$\bar{4}$ & T$_{2u}$ (I$\bar{4}$2d) & T$_{1g}$ (I4$_1$/a) & E$_u$ (I$\bar{4}$m2) \\
F222 & E$_u$ (I4$_1$22) & E$_g$ (Fddd) & E$_u$ (I$\bar{4}$m2) \\
\end{tabular}
\end{ruledtabular}
\label{table1}
\end{table}

This brings us to the one single crystal x-ray diffraction study \cite{single}.  In the cubic phase, one has Bragg peaks
at (H,0,0), with H=4n and (0,0,L) with L=4n.  In the tetragonal phase, new Bragg peaks appear at H=4n+2.  But whether new Bragg peaks
appear for (0,0,L) at L=4n+2 depends on the space group.  For  I$\bar{4}$m2, they are allowed.  But for I$\bar{4}$2d and I4$_1$/a, they
are not allowed.  Indeed, Ref.~\onlinecite{single} find no evidence for these new (0,0,L) Bragg peaks, giving support to a picture where
I$\bar{4}$2d and I4$_1$/a are primary and I$\bar{4}$m2 secondary.

To gain more insight, we turn to a discussion of the crystal structure of Cd$_2$Re$_2$O$_7$ (Fig.~2).  It is composed of ReO$_6$
octahedra (the Re ligands are O(1) ions) and Cd$_2$O chains (these are O(2) ions).  O(1) ions are displaced by all three of the above
mentioned distortions, Re and Cd ions by only T$_{2u}$ and E$_u$ distortions, and O(2) ions not at all.  Therefore, primary versus
secondary could be associated with particular atom types.  In that context, in most pyrochlores related to Cd$_2$Re$_2$O$_7$,
the primary distortion is associated with the chains.  This is clear in diffuse scattering studies of Cd$_2$Nb$_2$O$_7$ \cite{diffuse}.
Interestingly, in Pb$_2$Ir$_2$O$_7$, it is known that the primary distortion is in the Pb$_2$O chains, but that the SHG arises from
coupling of these distorted chains to the IrO$_6$ octahedra \cite{hirata}.  This can be understood from the fact that for the laser energy
used, one is accessing excitations of the Ir d levels (similar considerations apply to the SHG data for Cd$_2$Re$_2$O$_7$, where
one is probing excitations involving the Re d levels).

\begin{figure}
\includegraphics[width=0.75\columnwidth]{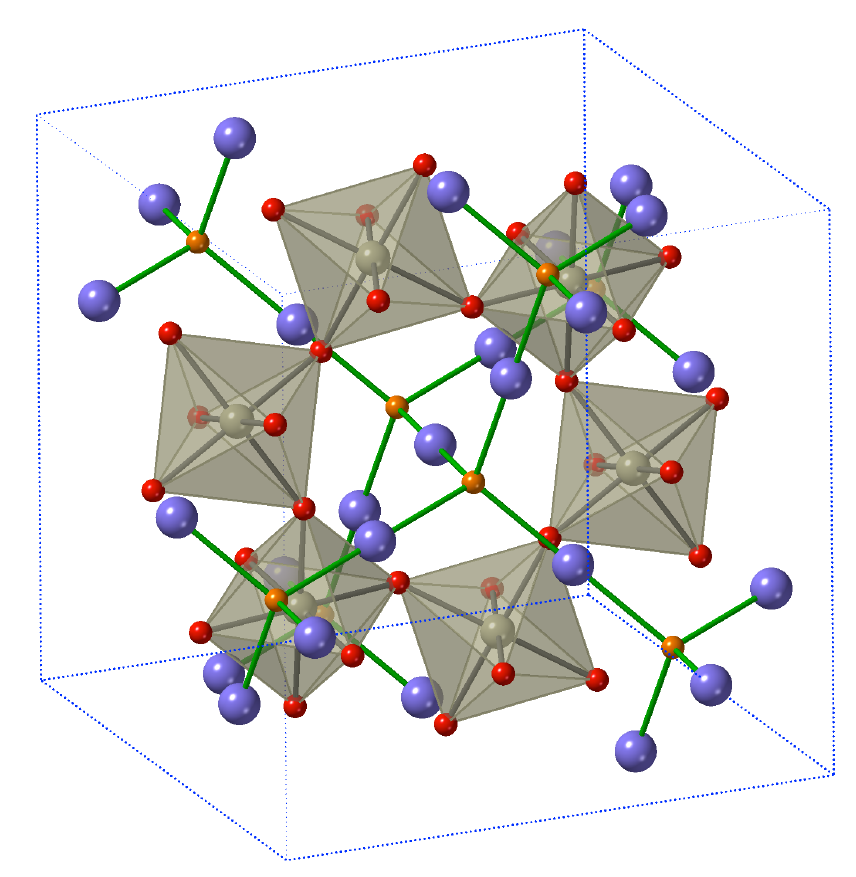}
\caption{Crystal structure of  Cd$_2$Re$_2$O$_7$.  Here, Re ions are in gray, Cd ions in purple, O(1) ions in red, and O(2) ions
in orange.  The crystal structure consists of corner sharing ReO$_6$ octahedra (the Re ions forming a pyrochlore lattice),
interpenetrated by Cd$_2$O chains.}
\label{fig2}
\end{figure}

This can be looked into in more detail by analyzing the two studies, one x-ray \cite{xray} and one neutron \cite{neutron}, that attempted
to determine atom displacements (a challenge given the weak nature of the distortions).  To do so, we analyze these using
AMPLIMODES \cite{amplimodes}.  The results are presented in Table II.  In both cases, the atom displacements assumed by these
authors was I$\bar{4}$m2.  This leads to a primary mode of E$_u$ symmetry, and secondary modes with symmetries E$_g$ and A$_{2u}$.
From Table II, one observes that both the mode amplitudes, and the individual atom type displacements, are significantly different
for the two refinements.  From x-rays, the ratio of the A$_{2u}$ to E$_u$ amplitudes is 57\%, whereas it is only 21\% for the neutrons.
Moreover, the x-rays indicate a significant displacement of the Re ions that is not observed by neutrons.  This suggests a serious
need to do a more careful study of the atom displacements, preferably supplemented by diffuse scattering studies.  A challenge is the multi-domain nature of the samples, where the c-axis of the tetragonal phase can point in each of the three cubic directions.  The domain structure
is known to be rich, and also qualitatively changes across the lower ($T_{s2}$) transition \cite{domains}.
Now, density functional theory studies have been done \cite{sergienko2}, but these have focused primarily on the O(1) ion displacements.  It
was found that E$_u$ was unstable and T$_{1u}$ not, but no other modes were commented on.
We note that Raman data have claimed that the transition at $T_{s2}$ is being driven by Cd ordering \cite{knee}.

\begin{table}
\caption{Mode amplitudes (in $\AA$) and the relative weight by atom type \cite{amplimodes} for Cd$_2$Re$_2$O$_7$ based on x-ray
diffraction at 160 K \cite{xray} and neutron diffraction at 13 K \cite{neutron}, relative to the high temperature cubic phase, assuming an I$\bar{4}$m2 space group.
For A$_{1g}$ and E$_g$, only the O(1) ions displace, with mode amplitudes of 0.0230 and 0.0326 $\AA$~from the x-rays,
and 0.0171 and 0.0584 $\AA$~from the neutrons.  The O(2) ions do not displace relative to their cubic positions.}
\begin{ruledtabular}
\begin{tabular}{llrrrr}
type & mode & Re & Cd & O(1) & total \\
\hline
x-ray & A$_{2u}$ & 0.060 & 0.341 & 0.599 & 0.0404 \\
x-ray & E$_u$ & 0.097 & 0.043 & 0.860 & 0.1882 \\
neutron & A$_{2u}$ & 0.002 & 0.132 & 0.866 & 0.1948 \\
neutron & E$_u$ & 0.000 & 0.363 & 0.637 & 0.3410 \\
\end{tabular}
\end{ruledtabular}
\label{table2}
\end{table}

\begin{figure*}
\includegraphics[width=2\columnwidth]{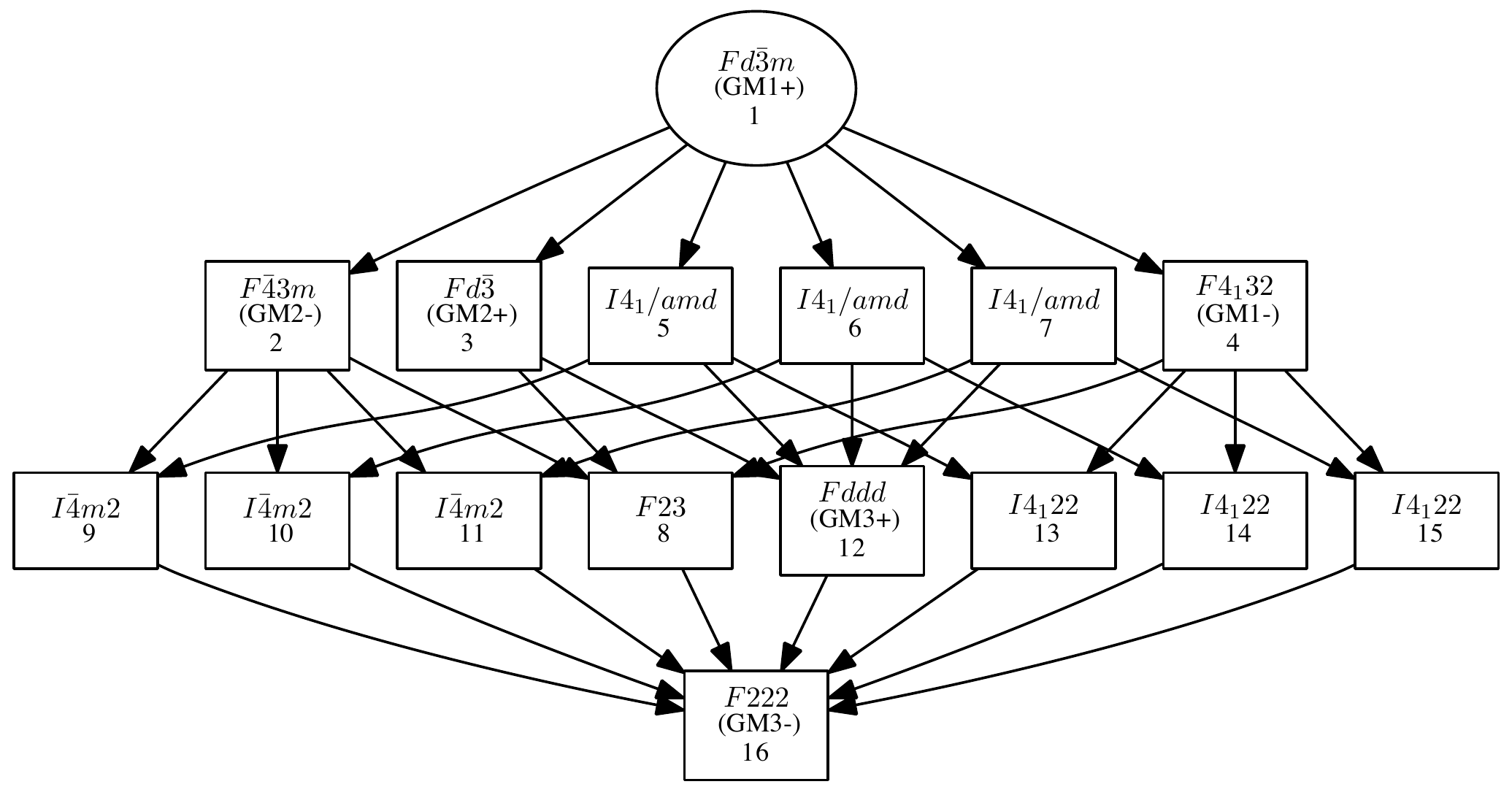}
\caption{Group-subgroup relations leading from Fd$\bar{3}$m to F222 \cite{bilbao}.  Here, the additional cubic group representations
not listed in the Fig.~1 caption are A$_{2g}$ ($\Gamma_2^+$, GM2+) and A$_{1u}$ ($\Gamma_1^-$, GM1-).}
\label{fig3}
\end{figure*}

This brings up the question whether other structural scenarios are possible.  The only other one that could be found is one
with an F222 space group, noting that a general combination of I$\bar{4}$m2 and I4$_1$22 forms this space group.   On the other
hand, a general F222 solution is not allowed \cite{sergienko} unless one couples E$_u$ to other representations.  The resulting
group-subgroup relation is significantly more complex (Fig.~3).  In this case, the three order parameters defining the trilinear term
in the Landau free energy differ from the I$\bar{4}$
scenario as summarized in Table I.  The primary order parameters are the second component of E$_u$ (i.e., I4$_1$22) and the second component of E$_g$
(the first is from I4$_1$/amd which also feeds into I$\bar{4}$, the second arises from Fddd), with again the first component
of E$_u$ being the secondary order parameter from I$\bar{4}$m2 (as before).
E$_u$(2) has the right angular symmetry to describe the primary order
parameter seen by SHG \cite{harter}.  And it also does not exhibit extra (00L) Bragg peaks,
consistent with Ref.~\onlinecite{single}.  As for E$_g$(2),
note that it is a secondary mode in the I$\bar{4}$m2 scenario \cite{sergienko} as shown in Table II.
In general, available diffraction data indicates a body centered tetragonal space group which F222 is not.
Moreover, a transition to F222 should be first-order in nature \cite{sergienko}.
Altogether, we find this F222 scenario to be less likely than the I$\bar{4}$ one.
Recently, though, Raman data have found evidence for a transition to F222 below 80 K \cite{karcia}.

In both structural cenarios, though, there are issues.  The most obvious one is why the primary order parameters are structural in nature despite
the small atom displacements, which is related to why the SHG signal is dominated by the secondary order parameter (which most likely is
structural in origin).
This is less of an issue in `electronic' scenarios where the primary order parameters have a non-structural origin \cite{fu,harter,matteo}.  On the
other hand, the SHG signal will be dominated by the Re d electrons, so the real issue is how the structural distortions couple to them.
Without a microscopic model, this is not an easy question to address.

The more serious issue, though, applies to all scenarios for Cd$_2$Re$_2$O$_7$.  In the electronic scenarios advocated by
Fu \cite{fu}, Harter {\it et al.}~\cite{harter}, and Di Matteo and Norman \cite{matteo}, and in both structural scenarios described here, the relevant 
term in the Landau free energy expansion is a
trilinear one involving three order parameters (two primary, one secondary), each from a different group representation.
It is hard to understand how two primary order parameters from different group representations can not only
simultaneously condense, but in a second-order fashion as experimentally observed, since this violates the Landau theory of phase transitions.
A similar issue is found in improper ferroelectrics which have a very similar trilinear term \cite{bousquet}.
Although there has been significant theory work on the latter \cite{benedek}, the question raised above is still unresolved for
these materials as well.
Therefore, a solution to this problem could
not only tell us much about Cd$_2$Re$_2$O$_7$, but about other materials where trilinear terms in the Landau free energy
expansion also play a fundamental role \cite{etxebarria}.
Finally, resolving the nature of the phase transition at $T_{s1}$ could potentially help us understand the origin of superconductivity
in Cd$_2$Re$_2$O$_7$.

In conclusion, the relative merits of a structural scenario for the SHG data of Cd$_2$Re$_2$O$_7$ have been presented.  From this,
it is clear that single crystal studies of the structure are needed to determine what its actual space group is, and also the
displacements of the various ions from their cubic positions in order to ascertain their potential impact on the nature of the SHG signal
(x-rays being more sensitive to the heavier Cd and Re ions, and neutrons requiring isotropically enriched Cd to mitigate absorption).
This could be aided by forcing the sample into a single domain by application of a magnetic field, given the anisotropy
seen in the magnetic susceptibility below $T_{s1}$ \cite{domains}.

This work was supported by the Materials Sciences and Engineering
Division, Basic Energy Sciences, Office of Science, U S Dept.~of Energy.


\begin{thebibliography}{99}

\bibitem{review}
For a recent review, see Z. Hiroi, J.-i. Yamaura, T. C. Kobayashi, Y. Matsubayashi and D. Hirai, Phys. Soc. Japan {\bf 87}, 024702 (2018).
\bibitem{hiroi}
J.-I. Yamaura and Z. Hiroi, J. Phys. Soc. Japan {\bf 71}, 2598 (2002).
\bibitem{xray}
S.-W. Huang, H.-T. Jeng, J.-Y. Lin, W. J. Chang,
J. M. Chen, G. H. Lee, H. Berger, H. D. Yang and K. S. Liang, J. Phys.: Condens. Matter {\bf 21}, 195602 (2009).
\bibitem{neutron}
M. T. Weller, R. W. Hughes, J. Rooke, C. S. Knee and J. Reading, Dalton Trans. {\bf 2004}, 3032 (2004).
\bibitem{single}
J. P. Castellan, B. D. Gaulin, J. van Duijn, M. J. Lewis, M. D. Lumsden, R. Jin, J. He, S. E. Nagler and D. Mandrus, Phys. Rev. B {\bf 66}, 134528 (2002).
\bibitem{petersen}
J. C. Petersen, M. D. Caswell, J. S. Dodge, I. A. Sergienko, J. He, R. Jin and D. Mandrus, Nature Phys. {\bf 2}, 605 (2006).
\bibitem{fu}
L. Fu, Phys. Rev. Lett. {\bf 115}, 026401 (2015).
\bibitem{harter}
J. W. Harter, Z. Y. Zhao, J.-Q. Yan, D. G. Mandrus and D. Hsieh, Science {\bf 356}, 295 (2017).
\bibitem{matteo}
S. Di Matteo and M. R. Norman, Phys. Rev. B {\bf 96}, 115156 (2017).
\bibitem{japan}
S. Hayami, Y. Yanagi, H, Kusunose and Y. Motome, Phys. Rev. Lett. {\bf 122}, 147602 (2019).
\bibitem{yamaura}
J. Yamaura, K. Ohgushi, H. Ohsumi, T. Hasegawa, I. Yamauchi, K. Sugimoto, S. Takeshita, A. Tokuda, M. Takata, M. Udagawa, M.
Takigawa, H. Harima, T. Arima and Z. Hiroi, Phys. Rev. Lett. {\bf 108}, 247205 (2012).
\bibitem{malcherek}
T. Malcherek, U. Bismayer and C. Paulmann, J. Phys.: Condens. Matter {\bf 22}, 205401 (2010).
\bibitem{bousquet}
E. Bousquet, M. Dawber, N. Stucki, C. Lichtensteiger, P. Hermet, S. Gariglio, J.-M. Triscone and P. Ghosez, Nature {\bf 452}, 732 (2008).
\bibitem{bilbao}
http://www.cryst.ehu.es/
\bibitem{isotropy}
http://iso.byu.edu/iso/isotropy.php
\bibitem{diffuse}
M. Pasciak, M. Wolcyrz, A. Pietraszko and S. Leoni, Phys. Rev. B {\bf 81}, 014107 (2010).
\bibitem{hirata}
Y. Hirata, M. Nakajima, Y. Nomura, H. Tajima, Y. Matsushita, K. Asoh,
Y. Kiuchi, A. G. Eguiluz, R. Arita, T. Suemoto and K. Ohgushi, Phys. Rev. Lett. {\bf 110}, 187402 (2013).
\bibitem{amplimodes}
D. Orobengoa, C. Capillas, M. I. Aroyo and J. M. Perez-Mato, J. Appl. Cryst. {\bf 42}, 820 (2009).
\bibitem{domains}
Y. Matsubayashi, D. Hirai, M. Tokunaga and Z. Hiroi, J. Phys. Soc. Japan {\bf 87}, 104604 (2018).
\bibitem{sergienko2}
I. A. Sergienko, V. Keppens, M. McGuire, R. Jin, J. He, S. H. Curnoe, B. C. Sales, P. Blaha, D. J. Singh,
K. Schwarz and D. Mandrus, Phys. Rev. Lett. {\bf 92}, 065501 (2004).
\bibitem{knee}
C. S. Knee, J. Holmlund, J. Andreasson, M. Kall, S. G. Eriksson and L. Borjesson, Phys. Rev. B {\bf 71}, 214518 (2005).
\bibitem{sergienko}
I. A. Sergienko and S. H. Curnoe, J. Phys. Soc. Japan {\bf 72}, 1607 (2003).
\bibitem{karcia}
K. J. Kapcia, M. Reedyk, M. Hajialamdari, A. Ptok, P. Piekarz, F. S. Razavi, A. M. Oles and R. K. Kremer,
arXiv:1911.11057.
\bibitem{benedek}
N. A. Benedek and C. J. Fennie, Phys. Rev. Lett. {\bf 106}, 107204 (2011).
\bibitem{etxebarria}
I. Etxebarria, J. M. Perez-Mato and P. Boullay, Ferroelectrics {\bf 401}, 17 (2010).

\end{thebibliography}
\end{document}